\documentstyle[aps,prl,twocolumn,epsfig]{revtex}

\renewcommand{\P}{\partial}
\renewcommand{\phi}{\varphi}
\newcommand{\pt}{p_{\rm T}}

\title{\large \bf Anisotropic flow from AGS to LHC energies}   
\author{P.~F.~Kolb$^a$ \and J.~Sollfrank$^a$ \and U.~Heinz$^{b,}$\thanks{On 
        leave of absence from Institut f\"ur Theoretische Physik, 
        Universit\"at Regensburg. Email: Ulrich.Heinz@cern.ch}}
\address{$^a$Institut f\"ur Theoretische Physik, Universit\"at
         Regensburg, D-93040 Regensburg, Germany\\
         $^b$Theoretical Physics Division, CERN, CH-1211 Geneva 23, 
         Switzerland}
\date{\today}

\begin{document}
%
%
\maketitle
\begin{abstract}
Within hydrodynamics we study the effects of the initial spatial 
anisotropy in non-central heavy-ion collisions on the momentum 
distributions of the emitted hadrons. We show that the elliptic 
flow measured at midrapidity in 158 $A$ GeV/$c$ Pb+Pb collisions 
can be quantitatively reproduced by hydrodynamic expansion, indicating 
early thermalization in the collision. We predict the excitation functions
of the 2$^{\rm nd}$ and 4$^{\rm th}$ harmonic flow coefficients from AGS 
to LHC energies and discuss their sensitivity to the quark-hadron phase 
transition.
\end{abstract}

\vskip 3truemm
%
The recent observation of transverse collective flow phenomena in 
non-central heavy-ion collisions at ultrarelativistic beam energies
\cite{E877_94,NA49_2_98,WA98_98,CERES_98} has led to renewed intense 
theoretical interest in this topic (see \cite{Olli_98} for a review). 
Collective flow is the consequence of pressure in the system and 
thereby provides access to the {\em equation of state} of the hot and 
dense matter (``fireball'') formed in the reaction zone. This access 
is indirect since the flow in the final state represents a time integral 
over the pressure history of the fireball. Sorge \cite{Sorge97} has 
ar\-gued that different types of transverse flow (radial, directed, 
elliptic, see \cite{Olli_98}) show different sensitivities to the early 
and late stages of the collision such that a combination of flow 
observables may allow for a more differential investigation of the 
equation of state. In particular, he pointed out that the {\em elliptic 
flow} (which develops in non-central collisions predominantly at 
midrapidity and manifests itself as an elliptic deformation of the
hadronic momentum distributions around the beam axis \cite{Stoe_82,O92}) 
is a signature for the {\em early stage} of the collision: its driving 
force is the spatial eccentricity of the dense nuclear overlap region 
which, if thermalized quickly enough, leads to an anisotropy of the 
pressure gradients which cause the expansion. Since the developing 
anisotropic flow reduces the eccentricity of the fireball, it acts 
against its own cause and thus shuts itself off after some time. Radial 
flow, on the other hand, responds to the absolute magnitude of the 
pressure gradients and not only to their anisotropy; it therefore 
exists also in central collisions, and in non-central collisions it 
continues to grow even after the initial elliptic spatial deformation 
of the fireball has disappeared.   

A phase transition from a hadron gas to a color-deconfined quark-gluon 
plasma causes a softening of the equation of state: as the temperature 
crosses the critical value for the phase transition, the energy and 
entropy densities increase rapidly while the pressure rises slowly.
The resulting small ratio of $p/e$ at the upper end of the 
transition region (``the softest point'' \cite{HS95}) weakens the 
build-up of flow as the system passes through it. Shuryak \cite{SZ78} 
and van Hove \cite{vH82} therefore suggested that a plot of the mean
transverse momentum against the central multiplicity density should 
show a plateau. Later hydrodynamic calculations did not confirm the 
existence of a plateau, showing only a slight flattening of an otherwise 
strictly monotonic curve \cite{KRMG86}. While the acceleration of the 
matter is weak in the transition region, the system also takes a long 
time to cross it, thereby allowing for the flow to build up over a 
longer time. This considerably reduces the sensitivity of the final 
radial flow to the existence of a soft region in the equation of state.

Recently van Hove's idea was revived in connection with elliptic flow
\cite{HL99,Sorge99}. Sorge \cite{Sorge99} used a modified version of 
RQMD which allows to simulate an equation of 
state with a ``softest point'' and found that the response of the 
elliptic flow to the initial fireball eccentricity was weakened for 
initial conditions in the phase transition region. Using a hydrodynamic
model, Teaney and Shuryak \cite{TS99} argued that the existence of the 
phase transition should, at higher energies, also lead to other dramatic 
effects in the transverse expansion pattern of non-central collisions, 
in particular to the formation of two well-separated shells moving into 
the reaction plane. In the present Letter we follow up on these ideas, 
trying to understand in more detail the transverse dyna\-mics in 
non-central collisions and what experimental data can tell us about it. 
We use a similar hydrodynamic approach as in \cite{TS99,O92}, adjust its 
free parameters to data from central Pb+Pb collisions at the SPS, 
demonstrate that it correctly reproduces the measured elliptic flow of
pions and protons at midrapidity \cite{NA49_2_98,NA49_WEB}, and then use 
it to make predictions at other beam energies. In particular we discuss 
the sensitivity of the excitation functions of $v_2$ and $v_4$, the 
2$^{\rm nd}$ and 4$^{\rm th}$ harmonic flow coefficients, to the existence 
of a deconfining phase transition.
   
In the hydrodynamic model one assumes that shortly after the impact
the produced strongly interacting matter reaches a state of local 
thermal equilibrium and subsequently expands adiabatically. In the 
conservation laws for energy-momentum and baryon number
 \begin{equation}
 \label{eq1}
   \P_\mu T^{\mu \nu}(x)=0\,, \quad 
   \P_\mu j^{\mu}(x)=0
 \end{equation}
one can then use the ideal fluid decompositions 
$T^{\mu \nu}= (e+p)u^\mu u^\nu-g^{\mu\nu}p$, $j^{\mu}=n\, u^{\mu}$ 
in terms of the energy density $e$, the pressure $p$, the (net) baryon 
number density $n$, and the fluid four-velocity $u^{\mu}$. One thus 
obtains the equations of ideal (non-dissipative) relativistic 
hydrodynamics. An equation of state $p(e,n)$ is needed to close the set 
of equations; its direct connection with the developing flow pattern 
makes hydrodynamics the most appropriate framework for an investigation 
of the equation of state.

We are here mostly interested in the transverse expansion dynamics in 
non-central ($b\ne 0$) heavy-ion collisions. The lack of azimuthal 
symmetry leads to a non-trivial 3+1 dimensional problem, requiring 
considerable numerical resources. As noted in \cite{O92}, the 
complexity of the task is significantly reduced if one focusses 
on the transverse plane at midrapidity and assumes that the 
longitudinal expansion can be described analytically by Bjorken's 
scaling solution \cite{B83} $v_z=z/t$. The latter is known to 
correctly reproduce the longitudinal expansion dynamics at 
asymptotically high beam energies, and it works phenomenologically very 
well even at SPS and AGS energies \cite{D99}. This assumption reduces 
the numerical problem to 2+1 dimensions. While it should be harmless 
at midrapidity, it forbids to make reliable predictions at forward and 
backward rapidities. Hence we cannot describe the rapidity dependence 
of the transverse flow pattern.

We investigated three different equations of state: (i) an ideal gas
of massless particles, $p={e\over 3}$ (EOS~I); (ii) a hadron resonance 
gas including all known resonances \cite{ParDat} with masses below 2 GeV 
and a repulsive mean field potential ${\mathcal V}(n)=\frac{1}{2} K n^2$,
with $ K=0.45$ GeV\,fm$^3$ \cite{Sol1_97} (EOS~H; for small $n$ this 
equation of state can be well characterized by the simple relation 
$p=0.15\,e$); (iii) an equation of state with a first order phase 
transition at $T_{\rm c}(n{=}0) = 164$ MeV, constructed by 
matching EOS~H and EOS~I using a bag constant $B^{1/4}= 230$ MeV 
(EOS~Q) \cite{Sol1_97}. EOS~Q features at $n=0$ a latent heat of 
1.15 GeV/fm$^3$: the mixed phase ranges from $e_{\rm H} = 0.45$ 
GeV/fm$^3$ to $e_{\rm Q} = 1.6$ GeV/fm$^3$. We show results only for 
the semi-realistic cases EOS~H and EOS~Q.

For $b\ne 0$ the initial energy density distribution in the transverse 
plane has an almond shape, characterized by an {\em eccentricity} 
$\epsilon_x = {\langle\!\langle y^2 - x^2\rangle\!\rangle 
\over \langle\!\langle y^2 + x^2\rangle\!\rangle} > 0$. 
($x$ denotes the transverse direction parallel to the impact 
parameter $\bbox{b}$, and $\langle\!\langle \dots \rangle\!\rangle$ 
indicates an energy density weighted spatial average.) This results 
in larger pressure gradients and thus in larger flow velocities in 
$x$ than in $y$ direction. Hence the final $\pt$-distribution 
is anisotropic. Its azimuthal angular dependence can be 
characterized by (even) Fourier coefficients \cite{VZ96}
$v_2,v_4,\dots$ (at midrapidity the odd ones, in particular the 
``directed flow'' $v_1$, vanish by symmetry):
 \begin{eqnarray}
 \label{eq2}
   &&{dN\over d\tilde y d\phi} = {dN\over 2\pi d\tilde y}
   \bigl(1{+}2\,v_2 \cos(2\phi){+}2\,v_4\cos(4\phi)+\dots\bigr),
 \\
 \label{eq3}
   &&{dN\over d\tilde y \pt d\pt d\phi} = 
   {dN\over 2\pi d\tilde y  \pt d\pt}
   \bigl(1 + 2\,v_2(\pt) \cos(2\phi) 
 \nonumber\\
   &&\qquad\qquad\qquad\qquad\qquad\quad
           + 2\,v_4(\pt)\cos(4\phi)+\dots\bigr)\,.
 \end{eqnarray}
($\tilde{y}=\mbox{Artanh} (p_z/E)$ is the longitudinal rapidity of 
the particles, and $\tilde y_{\rm cm}$ below denotes the
midrapidity point.)

For each impact parameter $b$, we parametrize the initial transverse
energy density $e(\bbox{r})$ by a Glauber-inspired formula which 
assumes that the deposited energy is proportional to the number of 
collisions producing wounded nucleons (for details see \cite{O92,Kol_99}). 
The initial baryon number density $n(\bbox{r})$ is taken proportional 
to $e(\bbox{r})$. The proportionality factors 
(in particular their $\sqrt{s}$-dependence) cannot be calculated
but must be adjusted to data. For adiabatic hydrodynamic expansion 
there exists a unique relation between the initial entropy density 
(related to $e_0,n_0$ by the equation of state) and the final 
multiplicity density $dN/d\tilde y$. We therefore present excitation 
functions as functions of the (total) pion multiplicity density at 
midrapidity, where the ``energy calibration'' 
$dN_\pi/d\tilde y \vert_{\tilde y=\tilde{y}_{\rm cm}}(\sqrt{s})$ 
will be provided by experiment.

The final particle distributions $dN_i/(d\tilde{y} \pt d\pt d\phi)$ 
are calculated using the Cooper-Frye formula \cite{CF74} with a 
freeze-out hypersurface of constant temperature. We adjust the 
model parameters, i.e. the initial central energy and baryon 
densities at $b$=0, $e_0$ and $n_0$, the equilibration time $\tau_0$, 
and the decoupling temperature $T_{\rm dec}$, by fitting \cite{Kol_99} 
the measured \cite{NA49_99} negative hadron and proton spectra from 
central 158 $A$\,GeV Pb+Pb collisions. For EOS~Q we find 
$T_{\rm dec}$=120 MeV, $e_0$=9.0 GeV/fm$^3$, $n_0$=0.95 fm$^{-3}$, 
and $\tau_0$=0.8 fm/$c$. The freeze-out temperature and the average 
radial flow resulting from these initial conditions agree well with 
previous studies \cite{Kaem_96,NA49_HBT,Tom_99}. In calculating the 
negative hadron spectrum we included \cite{Sol_90} decays of all 
resonances up to the mass of the $\Delta(1232)$; resonance decays 
are found to reduce the momentum anisotropies $v_{2,4}$ for pions 
by 10-15\%.

\begin{figure}[htbp]
  \begin{center}
  \epsfxsize 80mm 
  \epsfbox{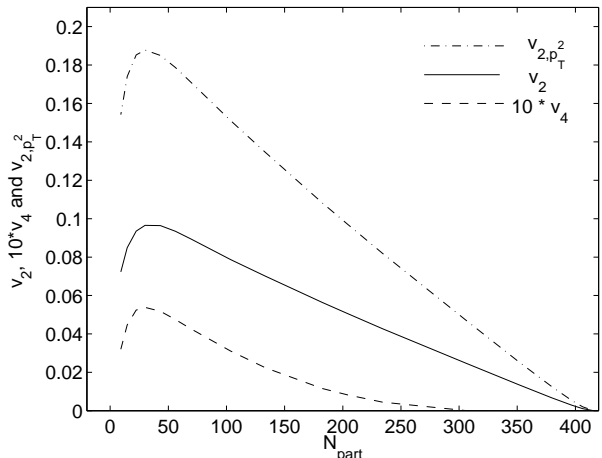}
    \caption{The 2$^{\rm nd}$ and 4$^{\rm th}$ harmonic flow coefficients 
      for pions as functions of the number of participating nucleons. 
      Also shown is the momentum-space anisotropy $\epsilon_p$ (see
      text). }
    \label{F1}
  \end{center}
\end{figure}

Having adjusted the model parameters in $b$=0 collisions, we can
calculate the initial density distributions also for $b\ne 0$ 
collisions, using the same Glauber formula. The equilibration time 
$\tau_0$ and decoupling temperature $T_{\rm dec}$ are left unchanged.
We have tested this procedure on $\pt$-spectra from non-central 
Pb+Pb(Au) collisions for pions \cite{WA98_98} and protons 
\cite{CERES_98,pbm98} and found very good agreement between 
data and hydrodynamic simulations up to impact parameters of about 
10 fm \cite{Kol_99}. We then proceed to compute the 2$^{\rm nd}$ and 
4$^{\rm th}$ harmonic coefficients $v_2$ and $v_4$ in Eq.~(\ref{eq2})
as functions of the number of participating nucleons, $N_{\rm part}$ 
(or, equivalently, as functions of the impact parameter $b$). The
results, for EOS~Q and the model parameters given above, are shown in 
Fig.~\ref{F1}.

In addition to $v_2=\langle \cos(2\phi)\rangle$ (where the average is 
taken with the particle momentum distribution) we also show the 
$\pt^2$-weighted elliptic flow (denoted as $\bar\alpha$ in \cite{O92})
 \begin{equation}
 \label{eq4}
   v_{2,\pt^2} = {\langle p_x^2 \rangle - \langle p_y^2 \rangle \over
                  \langle p_x^2 \rangle + \langle p_y^2 \rangle}
               = {\langle \pt^2 \cos(2\phi)\rangle \over
                  \langle \pt^2 \rangle}\, .
 \end{equation}
Fig.~\ref{F1} shows that for pions $v_2$ and $v_{2,\pt^2}$ differ by an 
overall factor of about 2, but otherwise have the same impact parameter 
dependence. 

\begin{figure}[htbp]
  \begin{center}
  \epsfxsize 80mm 
  \epsfbox{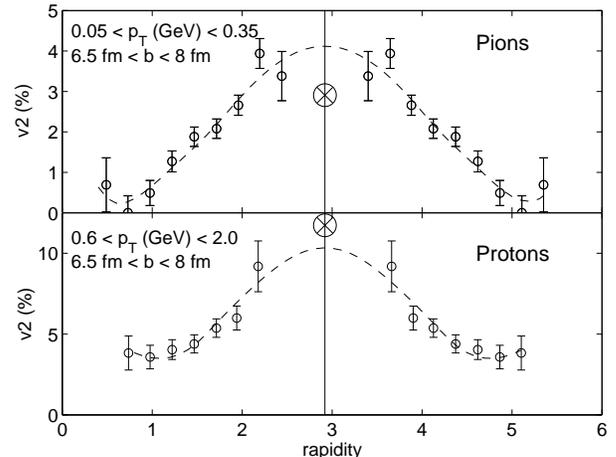}
    \caption{Elliptic flow $v_2$ for pions and protons, as a function
      of rapidity, as measured by NA49 in 158 $A$ GeV Pb+Pb collisions
      \protect\cite{NA49_WEB}. The dashed lines are to guide the eye.
      The circled crosses at midrapidity show our hydrodynamic
      results, with the same cuts in $b$ and $\pt$ as the data.}
    \label{F2}
  \end{center}
 \end{figure}

This factor 2 is important: previously $v_2$ and $v_{2,\pt^2}$ have 
often been used synonymously and, based on Ollitrault's results 
\cite{O92}, one concluded that hydrodynamic calculations overpredict 
the elliptic flow at the SPS by about a factor of 2. Fig.~\ref{F2} shows 
that this is not the case: a correct comparison of the data with the 
calculated $v_2$ (not $v_{2,\pt^2}$!) shows good quantitative 
agreement. The data \cite{NA49_WEB} were obtained from Pb+Pb collisions 
at the SPS, with a cut on the collision centrality and on the particle 
$\pt$ as given in the figure. Our calculation was done for $b$=7 fm and, 
using Eq.~(\ref{eq3}), the same $\pt$-cut as in the data was applied.
The resulting values for $v_2$ at midrapidity are 
$2.9\,\%$ for pions and $11.7\,\%$ for protons. The calculated 
$\pt$-dependence of $v_2$ (not shown) also agrees with the data 
\cite{Kol_99}, up to a normalization factor which takes into account 
that we compute $v_2(\pt)$ at midrapidity while the data \cite{NA49_2_98} 
were obtained in $4<\tilde y<5$ where $v_2$ is about a factor 3 smaller 
(see Fig.~\ref{F2}).

The good agreement of the data (the shape of the $\pt$-spectra
as a function of $b$ and the absolute values and $\pt$-dependences 
of $v_2$) with hydrodynamic calculations strongly suggests 
{\em very early thermalization and pressure buildup} in these collisions. 
In the calculation we can follow the time history of the elliptic flow: 
we found that $v_{2,\pt^2}$ for pions is nearly identical to
 \begin{equation}
 \label{eq5}
   \epsilon_p  
            = {\langle\!\langle T_{xx}-T_{yy} \rangle\!\rangle
               \over
               \langle\!\langle T_{xx}+T_{yy} \rangle\!\rangle}\,,
 \end{equation}
if the spatial average $\langle\!\langle \dots \rangle\!\rangle$ is 
performed at the time when the fireball center freezes out. $\epsilon_p$ 
is the momentum-space analog of the spatial eccentricity $\epsilon_x$
defined above; it does not require knowledge of the particle spectra 
and can be evaluated also at other times from the solution of the 
hydrodynamic equations. As expected we find that $\epsilon_p$ saturates 
as soon as the spatial anisotropy $\epsilon_x$ goes to zero. For Pb+Pb 
collisions at the SPS, $1\over 6$ of the final elliptic flow is created 
while the fireball center is in a pure QGP phase, ${1\over 2}$ of it is 
created in the mixed phase, and the final ${1\over 3}$ is generated during 
the hadronic stage. This agrees with Sorge's conclusion \cite{Sorge99} 
that the elliptic flow at the SPS indeed probes the deconfining phase 
transition and the existence of a QGP phase.

However, is it also {\em sensitive} to the existence of a phase 
transition? To answer this question we recalculated $v_2$ and $v_4$
with EOS~I and EOS~H, readjusting the initial conditions to the
measured $h^-$ and $p-\bar p$ spectra from central Pb+Pb collisions
\cite{Kol_99}. (While for EOS~H an acceptable fit is possible, the fit 
for EOS~I is quite bad, as found before by several other authors.) 
Whereas EOS~I (which can already be excluded from the $b=0$ spectra) 
gives about 30-40\,\% larger values for $v_2$, the elliptic flow 
developed by EOS~H is quite similar to that of EOS~Q. $v_4$ is about 
60\,\% larger with EOS~H than with EOS~Q. The time history
of $\epsilon_p$ reveals that the softening of the EOS near the 
phase transition delays the buildup of elliptic flow by about 
1.5-2 fm/$c$ but that, at this particular beam energy, in the end 
it reaches the same value. The mechanism is the same as discussed 
in the context of van Hove's plateau: the phase transition weakens 
the elliptic flow, but since the system also spends more time in the 
transition region, its net effect on $v_2$ is much less than naively 
expected.

\begin{figure}[htbp]
  \begin{center}
  \epsfxsize 80mm 
  \epsfbox{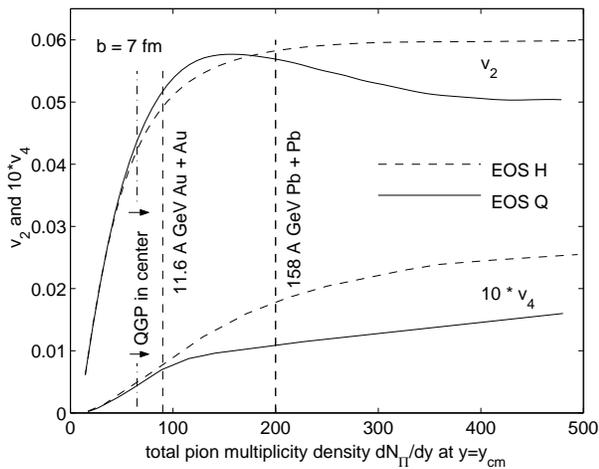}
    \caption{Hydrodynamic excitation functions of the 2$^{\rm nd}$ and 
      4$^{\rm th}$ harmonic flow coefficients, $v_2$ and $v_4$, for pions 
      from $A+A$ collisions ($A\approx 200$) at impact parameter $b$=7 fm. 
      The vertical dashed lines indicate the produced total pion 
      multiplicity densities at midrapidity for 11.6 and 158 $A$ GeV 
      beam energy (upper ends of the AGS and SPS ranges). The 
      dash-dotted vertical line indicates the threshold above which, 
      at $b$=7 fm, the fireball center is initially in a pure QGP phase.}
    \label{F3}
  \end{center}
\end{figure}

Given the apparent insensitivity of elliptic flow to the phase 
transition {\em at a fixed beam energy}, one may still hope for 
distinctive features in the excitation function of anisotropic flow 
\cite{D98}. In Refs. \cite{HL99,Sorge99} it was suggested that, for 
initial conditions around the ``softest point'', the response 
${\epsilon_p\over \epsilon_x} \approx 2\,{v_2^\pi\over \epsilon_x}$ should 
develop a plateau-like structure: as the beam energy or the collision 
centrality is increased and the initial energy density in the fireball 
center rises from subcritical to supercritical values, one should see
a weaker elliptic flow response in the transition region. To check 
this expectation we calculated excitation functions for $v_{2,4}$ for 
Pb+Pb collisions at fixed impact parameter $b=7$ fm (Fig.~\ref{F3}). 
Since we only varied $e_0$ (the initial central energy density 
in $b{=}0$ collisions), leaving the other model parameters ($\tau_0$, 
$e_0/n_0$, and $T_{\rm dec}$) unchanged, the curves in Fig.~\ref{F3} 
correspond to constant initial eccentricity $\epsilon_x=0.25$. Hence they 
give directly the $\sqrt{s}$-dependence (parametrized via 
$dN/d\tilde y\vert_{\tilde y_{\rm cm}}$) of the response of the 
anisotropic flow to $\epsilon_x$.

Fig.~\ref{F3} covers the range $1\leq e_0 \leq 25$ GeV/fm$^3$. Above 
SPS energies, $v_2$ and $v_4$ are seen to approach constant asymptotic 
values. This reflects the {\em saturation of the anisotropic flow before 
freeze-out} and its character as an early stage signature. $v_2$ saturates 
before $v_4$: first the dominant elliptic spatial deformation disappears, 
the smaller higher-order deformations are washed out later. At lower 
energies, $v_2$ and $v_4$ drop due to the decreasing fireball lifetime: 
the initial central temperature approaches $T_{\rm dec}$, causing 
decoupling before the anisotropic flow could fully develop. While this 
is a generic feature we must caution that quantitatively our results 
become unreliable in this region: We kept $T_{\rm dec}$ fixed although 
at lower beam energies freeze-out is known to occur at lower temperatures 
\cite{H96}; this gives more time for flow buildup, leading to larger 
values $v_2 > 0$. On the other hand, below 1-2 $A$ GeV/$c$ the elliptic 
flow at $\tilde y_{\rm cm}$ builds up before the spectator nucleons have 
moved out of the way (as we assume); this ``inertial confinement'' causes 
the elliptic flow to develop perpendicular to the reaction plane 
(``squeeze-out'' \cite{Stoe_82}, $v_2 < 0$) instead of in-plane as in 
our calculations. Experimentally this sign change of $v_2$ occurs near 
$E_{\rm beam}=4\,A$\,GeV \cite{Pink98}. Neither of these two phenomena 
is, however, directly related to the existence of a phase transition, and 
above AGS energies our results are not affected by them. 

Fig.~\ref{F3} shows that the expected weakening of the anisotropic 
flow due to the phase transition sets in between AGS and SPS energies:
comparing the curves for the hadron resonance gas equation of state
(EOS~H, dashed) with those for EOS~Q (solid), one sees that above SPS 
energies both $v_2$ and $v_4$ are reduced for EOS~Q. At high energies 
the anisotropic flow coefficients thus show a qualitatively similar 
dependence on the EOS as the radial flow \cite{KRMG86}. The reduction 
for $v_2$ is relatively small (for pions $v_2$ drops from 6\,\% to 
5\,\%); $v_4$ is reduced by about 40\,\% but, since $v_4$ is so small,
this is harder to measure. The effect on $v_2$ is also more 
characteristic, due to the ``bump'' in $v_2$ between AGS and SPS 
energies resulting from the interplay of the various effects discussed 
above; for $v_4$ the softening of the EOS cannot break the basic monotony 
of the excitation function.

Within hydrodynamics, the basic phase transition signature in the 
anisotropic flow is thus a {\em bump} (and not the conjectured 
\cite{HL99,Shu99} plateau and second rise) in the excitation function 
of $v_2$. Instead of changing $\sqrt{s}$, one can also vary the impact 
parameter at fixed $\sqrt{s}$ to change the produced multiplicity density 
$dN_\pi/d\tilde y \vert_{\tilde y=\tilde{y}_{\rm cm}}$. This provides 
for an alternate way to study the curves shown in Fig.~\ref{F3}, albeit 
not at fixed spatial ellipticity $\epsilon_x$. We found \cite{Kol_99} 
that in this case the elliptic flow response ${\epsilon_p\over\epsilon_x}
\approx 2 {v_2^\pi\over \epsilon_x}$ develops a similar bump \cite{fn0}. 
Unfortunately, for 158 $A$ GeV Pb+Pb collisions it lies at the limit of 
the accessible impact parameter range (near $b$=11 fm) where the 
hydrodynamic approach is expected to break down \cite{fn1}; for the 
planned SPS run at 40 $A$ GeV, however, the bump should be clearly 
visible in semicentral Pb+Pb collisions.

\begin{figure}[htbp]
  \begin{center}
  \epsfxsize 80mm 
  \epsfbox{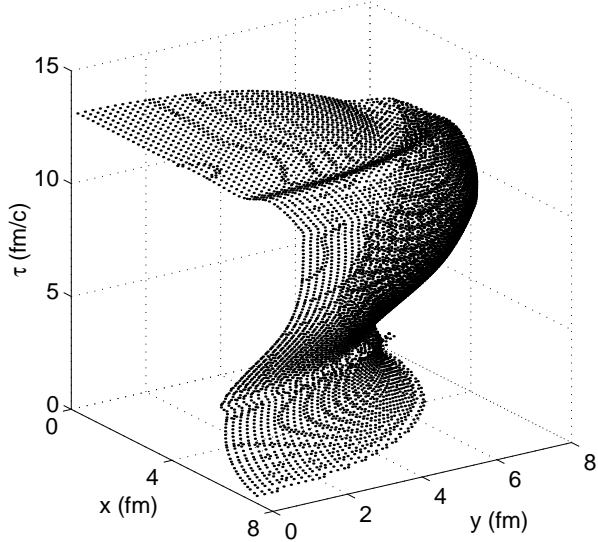}
    \caption{Freeze-out hypersurface $\tau(x,y,z{=}0)$ for $b{=}8$ fm 
      Pb+Pb collisions, for an initial temperature $T_0$=1.14 GeV
      ($dN_\pi/d\tilde y\vert_{\tilde y = \tilde y_{\rm cm}} \approx 1600$) 
      in central collisions.
      Note the dramatic transverse growths, followed by sudden
      freeze-out.}
    \label{F4}
  \end{center}
\end{figure}

As a last point we discuss why the seemingly so dramatic phenomenon
of the ``cracked nut'', recently advocated by Teaney and Shuryak 
\cite{TS99} as a hydrodynamic signature for the existence of a QGP 
phase transition, doesn't leave stronger traces in the anisotropic 
flow pattern. These authors argued that at high energies (RHIC or LHC)
a soft region in the EOS leads to the development of a ``shell'' at the 
edge of the almond-like initial fireball which is then cracked by the 
high pressure inside, with two separating half-shells expanding into 
the reaction plane. While we confirm their numerical results \cite{TS99}, 
we tend to interpret them more cautiously. To illustrate our point of 
view we show in Fig.~\ref{F4} the freeze-out surface $\tau(x,y)$ for a 
Pb+Pb collision at $b$=8 fm, initiated with a central temperature 
$T_0$=870 MeV at $\tau_0$=0.2 fm/$c$ and yielding a pion midrapidity density 
$dN_\pi/d\tilde y\vert_{\tilde y{=}\tilde y_{\rm cm}}{\approx}530$ 
(corresponding to 
$dN_\pi/d\tilde y\vert_{\tilde y{=}\tilde y_{\rm cm}}{\approx}1600$ 
for central collisions). Note its mushroom-like structure: While at 
the SPS hydrodynamics predicts freeze-out surfaces which shrink 
with time, the present surface features dramatic transverse growth 
\cite{KLRT92} before freezing out {\em nearly instantaneously} 
after 13 fm/$c$. One thus expects a very small emission time duration 
signal in two-particle Bose-Einstein correlations \cite{HBT},
in spite of the phase transition. 

Already before freeze-out the initial elliptic spatial deformation 
has vanished. The ripple on the top of the mushroom near its outer 
edge in $x$-direction is the ``nut shell'' \cite{TS99}: in a cut 
through the surface at $\tau\approx 13$ fm/$c$ it shows up as a 
crescent-shaped half shell at $x\sim 7$ fm. However, a mere 0.5 fm/$c$ 
later, the matter in this shell has frozen out, too. For EOS~H one 
finds a very similar mushroom, but without the ripple at the edge. 
Since there is no qualitative difference in the momentum-space 
structure of the ``shell'' compared to the rest of the matter, this 
explains why it is impossible to uniquely identify this particular 
structure by an anisotropic flow analysis. As suggested in \cite{TS99}, 
two-particle correlations may be more promising, but require extensive 
studies.  

We thank P.V.~Ruuskanen, M.~Kataja, R.~Venugopalan and P.~Huovinen for 
allowing us to use their hydrodynamics code for central collisions 
and to modify it for the study of non-central collisions. We are 
indebted to H.~Dobler, H. Heiselberg, A. Poskanzer and S. Voloshin for 
constructive remarks and acknowledge fruitful discussions with E. Shuryak 
and D. Teaney. This work was supported by BMBF, DFG and GSI.

\end{document}